\documentclass[12ptd, draftclsnofoot,onecolumn]{IEEEtran}
\usepackage{graphicx}
\usepackage{adjustbox}
\usepackage{setspace}
\usepackage{amsmath}
\usepackage{amssymb} 
\usepackage{bm} 
\usepackage{soul} 
\usepackage{dblfloatfix}
\usepackage{subcaption}
\usepackage{cite}
\usepackage{float}
\usepackage[usenames,dvipsnames]{pstricks}
\usepackage{epsfig}
\usepackage{pst-grad} 
\usepackage{pst-plot} 
\usepackage{array}
\allowdisplaybreaks

\newcommand{\nn} {\nonumber}

\begin{document}

\newcommand{\beq}{\begin{equation}}
\newcommand{\beqarr}{\begin{eqnarray}}
\newcommand{\beqarrn}{\begin{eqnarray*}}
\newcommand{\eeq}{\end{equation}}
\newcommand{\eeqarr}{\end{eqnarray}}
\newcommand{\eeqarrn}{\end{eqnarray*}}

\def\undb#1{\mbox{\bf{#1}}}


\title{Incremental Relaying for Power Line Communication: Performance Analysis and Power Allocation}

\author{ Ankit Dubey,~\IEEEmembership{Member,~IEEE,}
         Chinmoy Kundu,~\IEEEmembership{Member,~IEEE,}       
        Telex M. N. Ngatched,~\IEEEmembership{Senior Member, IEEE,} 
        Octavia A. Dobre, ~\IEEEmembership{Senior Member, IEEE,}
        and Ranjan K. Mallik, ~\IEEEmembership{Fellow, IEEE}
         
\thanks{
Manuscript received June 13, 2018; revised October 26, 2018; accepted October
28, 2018. 
This work was supported in part by the Department of
Science and Technology (DST), Govt. of India (Ref. No.
TMD/CERI/BEE/2016/059(G)), the Science and Engineering Research Board (SERB), Govt. of India through its Early Career Research (ECR) Award (Ref. No. ECR/2016/001377), Royal Society-SERB Newton
International Fellowship under Grant NF151345, and Natural
Science and Engineering Research Council of Canada (NSERC) through
its Discovery program.
(Corresponding author: Octavia
Dobre.)
}
        
      \thanks{Ankit Dubey is with the Department of ECE, National Institute of Technology Goa, Farmagudi, Ponda, Goa 403401, India, e-mail: ankit.dubey@nitgoa.ac.in.}  
        \thanks{Chinmoy Kundu is with the School of Electronics, Electrical Engineering and Computer Science, 
Queen's University Belfast, U.K.,  e-mail: c.kundu@qub.ac.uk.}

\thanks{Telex M. N. Ngatched and Octavia A. Dobre are with the Faculty of Engineering and Applied Science, Memorial University, Canada,
e-mail: tngatched@grenfell.mun.ca, odobre@mun.ca.}

\thanks{Ranjan K. Mallik is with the Department 
of Electrical Engineering, Indian Institute of Technology Delhi, New Delhi 110016, India, 
e-mail: rkmallik@ee.iitd.ernet.in.}

} 
\maketitle

\begin{abstract}
In this paper, incremental decode-and-forward (IDF) and incremental selective decode-and-forward (ISDF) 
relaying are proposed to improve the spectral efficiency of power line communication. Contrary to the 
traditional decode-and-forward (DF) relaying, IDF and ISDF strategies utilize the relay only if the 
direct link ceases to attain a certain information rate, thereby improving the spectral efficiency. 
The path gain through the power line is assumed to be log-normally distributed with high distance-dependent 
attenuation and the additive noise is from a Bernoulli-Gaussian process. Closed-form expressions for the 
outage probability, and approximate closed-form expressions for the end-to-end average channel capacity 
and the average bit error rate for binary phase-shift keying are derived. Furthermore, 
a closed-form expression for the fraction of times the relay is in use is derived as a measure of 
the spectral efficiency. Comparative analysis of IDF and ISDF with traditional DF relaying is presented. 
It is shown that IDF is a specific case of ISDF and can obtain optimal spectral efficiency without 
compromising the outage performance. By employing power allocation to minimize the outage probability, 
it is realized that the power should be allocated in accordance with the inter-node distances and channel parameters.
\end{abstract}


\IEEEpeerreviewmaketitle

\section{Introduction}
\label{sec_introduction}
Power line communication (PLC)
is a key solution that is driving Internet-of-Things (IoT) based
smart grid, home automation, and many such other concepts.
Being a retrofit technology,
PLC does not need extra communication links to be deployed,
unlike optical and other traditional wireline communication systems
\cite{PLC_BOOK:10}.
Furthermore, due to the omnipresence of power lines, unlike wireless system, PLC has 
greater device-to-device (appliance-to-appliance) connectivity that makes PLC one of 
the suitable candidates for IoT \cite{PLC_IoT_1:18,PLC_IoT_2:18_EA}.

Implementation cost and resource complexity for a PLC system
is much lower when compared to other wireline communication systems,
but it has its own hurdles.
A power line is designed to carry high power
alternating current signals at very low frequency (around $45$ to $65$ Hz);
on the contrary, in PLC systems, data symbols are
transmitted with very small signal power
over a high frequency carrier, and hence, are subjected to high signal attenuation \cite{PLC_BOOK:10}.
Additionally, due to the mesh structure of the power distribution,
the communication signal transmitted using power line
reaches the receiver
through multiple paths, which causes signal fading.
In many scenarios, fading in the PLC
channel is shown to follow the log-normal distribution
\cite{PaCaKaTh:03, GuCeAr:11, Ga:11}. 
Furthermore, due to frequent load switching and
integration of new home and industry appliances 
like light emitting diode lamps, stepper motors, air conditioners,
induction plates, etc.,
data symbols transmitted through power lines
suffer from impulsive noise \cite{PLC_BOOK:10, DuShMaMi:15}.

As in wireless communication, reliability of data transmission in PLC systems can also be
enhanced with the aid of relays \cite{LaScYi:06, LaVi:11}.
{Similar to a wireless system, the relay in a PLC system can be a dedicated or a cooperative node and can act in amplify-and-forward (AF) or decode-and-forward (DF) mode.}
 A study on cooperative coding for narrowband PLC
has been presented in \cite{LaVi:11}. 
A distributed space-time coding for multi-hop transmission has been introduced 
in \cite{LaScYi:06}.
A bound on channel capacity is derived using AF relays in \cite{ChCaYa:13}.
 An opportunistic routing for smart grid
with PLC access networks is presented in \cite{YoJaKiBa:14}.
Recently, the average bit error rate (BER) and outage analyses using DF and AF relays
have been studied in \cite{DuMaSc:15} and \cite{DuMa:15}, respectively, where the possibility of 
direct transmission has been overruled. 
Furthermore, to address the relay selection problem for dual-hop transmission, a class of machine learning schemes has been proposed in \cite{NiVi:17}. Recently, DF energy-harvesting based relaying has been proposed for PLC and the issue of energy-efficiency has also been discussed in \cite{RaAdToNa:17, RaAdGa:17}. {Very recently, the ergodic achievable data rate of the so-called incomplete hybrid power line-wireless single-relay channel model has been studied in} \cite{FePoRi:18}.

In the above literature, the relays are generally DF or AF type with half-duplex transmission capability, which requires two time slots to complete the 
information transmission from source to destination \cite{MH_CR_BOOK:08, Laneman_Wornell_cooperative_diversity}. {It is possible to improve the system spectral efficiency by using a relaying technique referred to as  incremental relaying} 
\cite{Laneman_Wornell_cooperative_diversity}. 
Incremental relaying exploits feedback from the destination to decide whether or not 
relaying is necessary for retransmission of the source message. In incremental relaying, the relay is used only if 
the direct transmission from source to destination fails to achieve a required information rate. 
In combination with DF relays, incremental relaying can be applied through the following two schemes:
i) incremental DF (IDF) or ii) incremental selective DF (ISDF). 
In the IDF technique, relaying takes place every 
time the direct transmission fails. On the other hand, in the ISDF strategy, the relay is utilized when the direct 
transmission fails and the source-to-relay link achieves a required information rate. 
Though incremental relaying has been investigated in wireless systems (see \cite{Hwang_incremental_relaying_with_relay_selection, 
ikki_Performance_analysis_of_incremental, Bai_Yuan_Performance_Analysis_of_SNR_Based} and the references therein), 
to the best of our knowledge, it has not been studied yet in depth in PLC systems. 
{Very recently, work reported in }\cite{our_conf_VTC:17} {has introduced incremental 
selective relaying for the PLC for the first time. However, contributions are limited 
to only independent and identically distributed (i.i.d.) channels. Moreover, the analysis 
is limited to the outage probability and the average BER.}

In this paper, the use of incremental relaying techniques, IDF and ISDF, is proposed and investigated
to enhance the spectral efficiency of PLC systems. 
It is assumed that the PLC channel suffers from log-normal fading with high distance-dependent attenuation, along with  
Bernoulli-Gaussian impulsive noise.
Closed-form expression for the outage probability and approximate closed-form expressions for the
end-to-end average channel capacity and average BER are derived and compared with traditional DF relaying. 
The fraction of times the relay is in use is also investigated as a measure of the spectral efficiency. 
Furthermore, a technique to minimize the outage probability with appropriate power allocated to the source and relay is investigated. 

\begin{figure}
[t]
\hspace{-1.5cm}
\begin{center}
\psfig{file=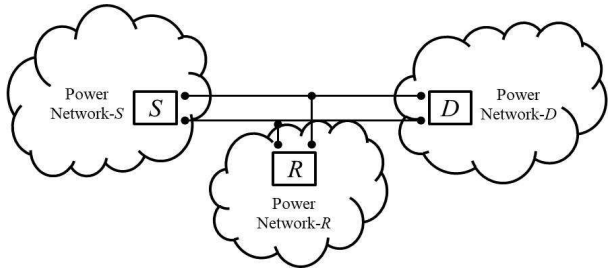,width=8cm,height=3.5cm}
\caption{{Relay based PLC system.}}
\label{fig_1}
\end{center}
\end{figure}  

The rest of the paper is organized as follows. Section \ref{sec_system} describes the system model, 
while an approximate closed-form expression for the end-to-end average channel capacity, a 
closed-form expression for the outage probability, and an approximate closed-form expression 
for the average BER are derived in Sections \ref{sec_capacity}, \ref{sec_outage}, and \ref{sec_ABER}, respectively. 
Later, a closed-form expression for the fraction of times the relay is in use is derived in Section 
\ref{relay_usgae}. The power allocation problem is investigated in Section \ref{sec_power}. 
Section \ref{sec_results} presents numerical and simulation results, whereas Section 
\ref{sec_conclusion} provides concluding remarks.

\textit{Notation:}
$\mathbb{E}[\cdot]$ denotes the expectation operator, 
$\mbox{Pr}(\cdot)$ denotes the probability of an event, $P_e(\cdot)$ denotes the probability of bit error,
$F_X (\cdot)$ represents the cumulative distribution function (CDF) of the random variable (r.v.) $X$, and 
$f_X (\cdot)$ is the corresponding probability density function (PDF). 

\section{System Model}
\label{sec_system}

\begin{figure}
\begin{center}
\psfig{file=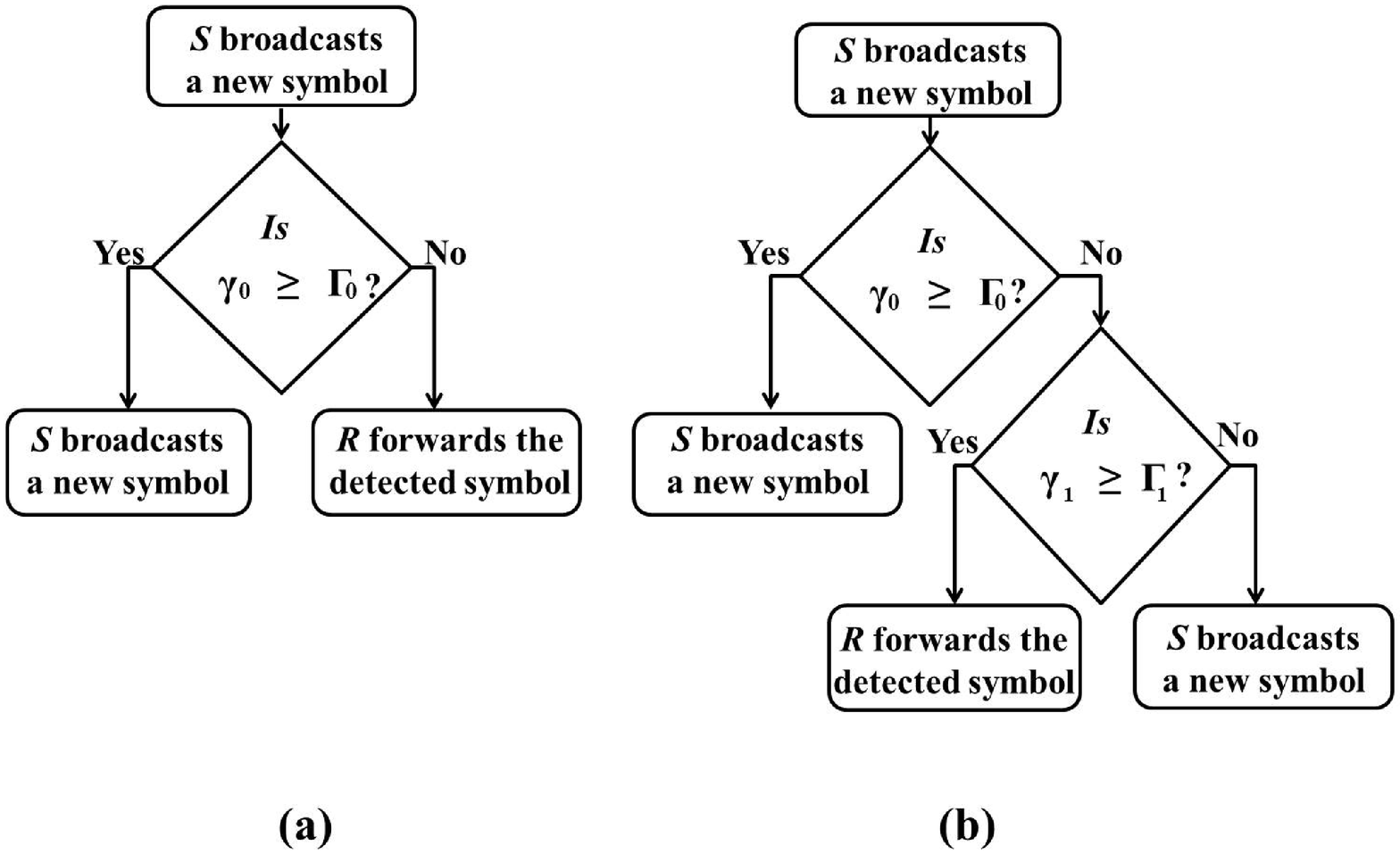,width=8.8cm,height=4.7cm}
\caption{Relaying strategies: (a) IDF and (b) ISDF.}
\label{fig_2}
\end{center}
\end{figure}

Consider a PLC system, 
{as shown in Fig.} \ref{fig_1}, 
wherein the source, $S$, communicates with the destination, $D$, over a power cable.
A half-duplex DF relay, $R$, is placed in between $S$ and $D$ to assist $S$ forward its symbols. 
A link between any two nodes is denoted by $i$, where 
$i\in\{0, 1, 2 \}$, 
and $0$, $1$, and $2$ denote the $SD$, $SR$, and $RD$ links, 
respectively. $S$ broadcasts a symbol with transmit power $P_S$ in the first time slot, which is received by $D$ and $R$. Two different relaying strategies, IDF and ISDF, are studied. 
In the IDF strategy, $R$ transmits the decoded symbol in the second time slot whenever the direct transmission ($SD$ link) fails to attain a predefined rate threshold, $R_{th}$, {or equivalently when the instantaneous signal-to-noise ratio (SNR) of the $SD$ link, $\gamma_0$, falls below a predefined threshold $\Gamma_0$}.
On the contrary, the ISDF strategy sets a predefined rate threshold {or equivalently an SNR threshold, $\Gamma_1$, at $R$ to restrict the transmission 
below a certain rate requirement through the $SR$ link, i.e., when the instantaneous SNR of the $SR$ link, $\gamma_1$, falls below $\Gamma_1$.}  
{Fig.} \ref{fig_2} {illustrates the comparison flowchart.} 
In both cases, the relayed link is only used if the direct transmission rate fails to attain $R_{th}$ to improve spectral efficiency. {Moreover, in both strategies, $S$ broadcasts a new symbol immediately after the preceding symbol if $R$ does not participate; otherwise, it waits until the relay completes forwarding the previous symbol.}

\subsection{Channel Model and Received Power}
The received symbol, $y_{i}$, through the $i$th link is given by
\beqarr
y_{i}= \sqrt{P_i} h_{i}s+z_{i},
\label{eq_revd}
\eeqarr
where $P_i$ is the received power, $h_{i}$ is the channel gain of the $i$th link, $z_{i}$ is the additive noise sample at the receiver, and $s$ is the unit power transmitted symbol. {Since all nodes have non-identical local power networks (nearby loads) attached to them, as shown in Fig. 1, the PLC channels among different nodes can be considered to be independent but non-identical} \cite{PaCaKaTh:03, GuCeAr:11, Ga:11}. 
{Thus, the channel gain multiplier, $h_i$,
is modeled as an independently distributed 
log-normal r.v. with PDF}
\beqarr
f_{h_{i}}(x) = \frac{1}{x\sqrt{2\pi \xi_i^2}}
\exp\left(-\frac{1}{2}\left(
\frac{\ln{x} - \Xi_{i}}{\xi_i}
\right)^2
\right) \, , \; \; x \geq 0,
\label{e12}
\eeqarr
where
the parameters $\Xi_{i}$ and $\xi_i$ are the mean 
and the standard deviation of the normal r.v. $\ln{(h_{i})}$, respectively.
The $\ell$th moment of $h_{i}$ is given as \cite{Ho:89}
\beqarr
\mathbb{E}\left[h_{i}^{\ell}\right] = \exp\left({\ell}\Xi_{i} +
\frac{{\ell}^2\xi_i^2}{2}\right) \,. 
\label{e13}
\eeqarr
We assume unit energy of the channel gain, i.e., $\mathbb{E}[h_{i}^2]=1$. According to (\ref{e13}), 
this implies 
$\Xi_{i}=-\xi_i^2 \,.$

The received power, $P_i$, depends on the transmit power, the length of the power cable, and the path loss. 
The total transmit power, $P_T$, is divided among $S$ and $R$
with fractions $p_f$ and $(1-p_f)$, respectively, where $0<p_f<1$.
Thus, the transmit powers of $S$ and $R$ can be expressed as
\begin{align}
&P_S=p_f P_T\,,
P_R=(1-p_f)P_T.
\label{e4}
\end{align}
The length of the power cable from $S$ to $D$, $S$ to $R$, and $R$ to $D$ are
$d_{0}$, $d_{1}$, and $d_{2}$, respectively.
The distances $d_{1}$ and $d_{2}$ can be expressed 
as
\begin{align}
&d_{1}=d_fd_{0},\,\,
d_{2}=(1-d_f) d_{0},
\label{e2}
\end{align}
where $0<d_f<1$. 
The dB equivalent of the received power through the $SD$, $P_{0}$, can be expressed as
\beqarr
P_{0} \mbox{(dB)}=P_{S} \mbox{(dB)}-d_{0}\mbox{(km)}\times P_L \mbox{(dB/km)}\,,
\label{e9}
\eeqarr
where $P_L\mbox{(dB/km)}$ denotes the path loss factor per unit length. 
The dB equivalent of $P_{1}$ and 
$P_{2}$ can also be similarly expressed.
\subsection{Noise Model and SNR}
The symbols transmitted through power lines
suffer from impulsive noise \cite{PLC_BOOK:10, DuShMaMi:15}.
Among several proposed noisy channel model, the 
Bernoulli-Gaussian model \cite{MaSoGu:05} 
is the mostly used \cite{DuMaSc:15},\cite{DuMa:15}.
The model consists of a mixture of two Gaussian
processes with different power spectral densities
and average occurrence of the impulsive noise
with Bernoulli process. 
According to this noise model, the additive noise sample, $z_{i}$,
can be written as
\beqarr
z_{i} = w_i + b_i l_i \,,
\label{e15}
\eeqarr
where $w_i$ and $l_i$
represent background and impulsive noise samples, respectively,
and $b_i$ is a Bernoulli r.v. which equals $1$ with probability $ \Lambda$ and $0$ with probability $(1- \Lambda)$. 
The samples $w_i$ and $l_i$
are taken from the Gaussian distribution with mean zero and variances
$\sigma_w^2$ and $\sigma_l^2$, respectively.
Furthermore, $w_i$, $l_i$, and $b_i$ are considered to be independent
as background and impulsive noise have different origins \cite{GoRaDo:04}. 
Therefore, the noise samples, $z_i$, are 
i.i.d. r.v.s, with PDF \cite{MaSoGu:05}
\beqarr
f_{z_{i}}(x) =\sum\limits_{j=1}^{2} \frac{ \Lambda_j}
{\sqrt{2\pi \epsilon_{j}^2}}\exp{\left(\frac{-x^2}
{2 \epsilon_{j}^2}\right)}
\, ,
\label{e16}
\eeqarr
where
\beq
 \Lambda_1=1- \Lambda \, , \; \;  \Lambda_2= \Lambda \, , \; \; \epsilon_{1}^2=\sigma_w^2 \, ,
\; \;
\epsilon_{2}^2=\sigma_w^2+\sigma_l^2 \, .
\label{e17}
\eeq
The average noise power, $N_{0}$, for all $i$
is given as
\beqarr
N_{0}=\mathbb{E}\left[ z_{i}^2 \right]
    = \mathbb{E}\left[ w_i^{2} \right]+\mathbb{E}\left[
    b_i^2 \right]\mathbb{E}\left[ l_i^{2} \right]
    = \sigma_w^2 (1+ \Lambda\ \eta) \, ,
\label{e18}
\eeqarr
where {
$\eta={\sigma_l^2}/{\sigma_w^2}$
}represents the power ratio of impulsive-to-background noise.

As the channel gain, $h_{i}$, is log-normally distributed, 
the corresponding instantaneous SNR, $\gamma_{i}=\frac{P_ih_{i}^2}{N_{0}}$, is also log-normally distributed with
 {parameters}
\beqarr
\mu_{i} = 2 \Xi_{i}+\ln{\left({P_{i}}/{N_{0}}\right)}\,,
\quad \sigma_{i} = 2\xi_i\,.
\label{e22}
\eeqarr
The CDF of $\gamma_i$ is therefore given by
\beqarr
F_{\gamma_{i}}(x)=
\mbox{Pr}[\gamma_{i}\leq x]=
1-
Q\left(
\frac{\ln{x}-\mu_{i}}{\sigma_{i}}
\right) \, , \; \; x \geq 0,
\label{e23}
\eeqarr
where $Q(\cdot)$ denotes the Gaussian $Q$-function defined as \cite{Si:02}
\begin{align}
Q(x)\triangleq \int_{x}^{\infty}\frac{1}{\sqrt{2\pi}}\exp\left(-\frac{t^2}{2}\right)\mbox{d}t.
\label{eq_qfunc}
\end{align} 

\subsection{SNR Threshold for Rate Requirement}

The instantaneous channel capacity of the $i${th} link corrupted by Bernoulli-Gaussian impulsive 
noise is expressed as \cite{WiStTu:09}
\beqarr
{\cal C}(\gamma_i)=
\sum_{j=1}^{2}
 \Lambda_j\log_2{\left(1+\tau_j\gamma_i\right)}\,\,,
\label{e24}
\eeqarr
where $\tau_1$ and $\tau_2$
are given by
\beqarr
\tau_1= \frac{1+ \Lambda\eta}{2},\quad\,
\tau_2= \frac{1+ \Lambda\eta}{2(1+\eta)}.
\label{e25}
\eeqarr
At high SNR ($P_i/N_0$), i.e., when $\tau_j\gamma_i>>1$,  the instantaneous channel capacity can be simplified as \cite{DuMaSc:15}
\begin{align}
{\cal C}(\gamma_i)\approx \frac{1}{\ln(2)}
\left(
\sum_{j=1}^{2}\ln(\tau_{j}^{ \Lambda_j})+
\ln(\gamma_i)
\right)\,.
\label{ec4}
\end{align}
To maintain $R_{th}$ at $D$ through the direct link, a specific SNR threshold, $\Gamma_{0}$, 
must be maintained; this can be evaluated from (\ref{ec4}) as 
\beqarr
\Gamma_{0} = {\tau_1}^{- \Lambda_1}{\tau_2}^{- \Lambda_2}2^{R_{th}}\,.
\label{e26}
\eeqarr 
To maintain $R_{th}$ at $D$ through the relayed link, the end-to-end SNR threshold in relayed link, $\Gamma_{th}$, should be twice the rate of the direct link due to the half-duplex relaying 
\begin{align}
\Gamma_{th}= {\tau_1}^{- \Lambda_1}{\tau_2}^{- \Lambda_2}2^{2 R_{th}}\,.
\label{e45a}
\end{align}

{If the relay is in operation, information takes two time slots to reach the destination. 
On the contrary, direct transmission only takes one time slot. The SNR threshold at the 
destination and relay determines how frequently the relay will be used, and thereby, 
the latency associated with it.}
Thus, in order to improve the spectral efficiency by reducing the relay usage, we 
selectively decode-and-forward the signal at the relay by setting an SNR threshold 
$\Gamma_{1}$ for the ISDF scheme. In a practical set up, the value of $\Gamma_{1}$ 
can be selected based on various factors like instantaneous battery level, minimum 
required rate, latency, and relay placement. In this work, we select a variable 
$\Gamma_1$ and evaluate the system performance.


\section{Average Channel Capacity}
\label{sec_capacity}
In this section, we derive the end-to-end average channel capacity for the relaying strategies. 
The average channel capacity is derived using the capacity definition provided in the high SNR in (\ref{ec4}) 
to alleviate the difficulty in dealing with (\ref{e24}) to find a closed-form. 
\subsection{IDF}
While employing IDF, the end-to-end average channel capacity can be calculated as the sum of 
average capacities of direct and relayed transmissions.
The average capacity of the direct transmission is calculated when the $SD$ link SNR meets the 
threshold requirement and in relayed transmission, it is calculated when the $SD$ link does not meet 
the threshold. Thus, the average capacity of the system can be obtained using the laws of 
total probability as 
\begin{align}
&{\cal C}_{\text{IDF}}
={\mathbb E}\left[{\cal C}(\gamma_{0}|\gamma_{0}>\Gamma_{0})\right]
+\frac{1}{2}\mbox{Pr}[\gamma_{0}<\Gamma_{0}]{\mathbb E}\left[{\cal C}(\gamma_m)\right]\,,
\label{ec1}
\end{align}
where $\gamma_{m}=\min\{\gamma_{1},\gamma_{2}\}$ and the 
factor $1/2$ in the second term comes from half-duplex relaying.
 {Since $\gamma_1$ and $\gamma_2$ are independent, the PDF of $\gamma_{m}$ can be expressed as}  
\begin{align}
\hspace*{-0.1cm}
f_{\gamma_{m}}(x)
&=
Q\left(
\frac{\ln{x}-\mu_{2}}{\sigma_{2}}
\right)
\hspace*{-0.1cm}
\frac{1}{x \sqrt{2\pi\sigma_{1}^2}}
\exp\left(
\hspace*{-0.1cm}
-\frac{1}{2}
\left(
\frac{\ln{x}-\mu_{1}}{\sigma_{1}}
\right)^2
\right)\nn\\
&+
Q\left(
\frac{\ln{x}-\mu_{1}}{\sigma_{1}}
\right)
\hspace*{-0.1cm}
\frac{1}{x \sqrt{2\pi\sigma_{2}^2}}
\exp\left(
\hspace*{-0.1cm}
-\frac{1}{2}
\left(
\frac{\ln{x}-\mu_{2}}{\sigma_{2}}
\right)^2
\right).
\label{ec3}
\end{align}
From (\ref{ec1}), (\ref{ec3}), and (\ref{ec4}), the end-to-end average channel capacity, for high SNR, can be expressed
in closed-form as
\begin{align}
{\cal C}_{\text{IDF}}
&
=\frac{1}{\ln(2)}
\left\{
\left(
\sum_{j=1}^{2}\ln(\tau_{j}^{ \Lambda_j}+\mu_{0})
\right)
Q\left(
\frac{\ln\Gamma_{0}-\mu_{0}}{\sigma_{0}}
\right)
\right.
\nn\\
&
+
\frac{\sigma_{0}}{\sqrt{2\pi}}
\exp\left(
-\frac{1}{2}
\left(
\frac{\ln\Gamma_{0}-\mu_{0}}{\sigma_{0}}
\right)^2
\right)
\nn\\
&
+
\frac{1}{2}
\left(
1-Q\left(
\frac{\ln\Gamma_{0}-\mu_{0}}{\sigma_{0}}
\right)
\right)
\left\lbrace
\left(
\sum_{j=1}^{2}\ln(\tau_{j}^{ \Lambda_j})
\right)
\right.
\nn\\
&
+
\mu_{1}Q\left(
\frac{\mu_{1}-\mu_{2}}{\sqrt{\sigma_{1}^2+\sigma_{2}^2}}
\right)
-\frac{\sigma_{1}^2}{\sqrt{2\pi(\sigma_{1}^2+\sigma_{2}^2)}}
\nn\\
&\times \exp\left(
-\frac{1}{2}
\left(
\frac{\mu_{1}-\mu_{2}}{\sqrt{\sigma_{1}^2+\sigma_{2}^2}}
\right)^2
\right)
+
\mu_{2}Q\left(
\frac{\mu_{2}-\mu_{1}}{\sqrt{\sigma_{2}^2+\sigma_{1}^2}}
\right)\nn\\
&-
\hspace*{-0.1cm}
\frac{\sigma_{2}^2}{\sqrt{\sigma_{2}^2+\sigma_{1}^2}}
\frac{1}{\sqrt{2\pi}}
\exp
\left.
\left.
\hspace*{-0.2cm}
\left(
\hspace*{-0.1cm}
-\frac{1}{2}
\left(
\frac{\mu_{2}-\mu_{1}}{\sqrt{\sigma_{2}^2+\sigma_{1}^2}}
\right)^2
\right)
\hspace*{-0.1cm}
\right\rbrace
\hspace*{-0.1cm}
\right\rbrace
.
\tag{21}
\label{ec8}
\end{align}

\begin{table*}[!b]
\begin{tabular}{m{\textwidth}}
\hrule{
\begin{align}
&S_n=\frac{\Omega_n\sigma_{2}}{\sqrt{\Omega_n^2 \sigma_{1}^2+2\sigma_{2}^2}}
\,,
T_n=\frac{\Omega_n\sigma_{1}}{\sqrt{\Omega_n^2 \sigma_{2}^2+2\sigma_{1}^2}}
\,,
E_n=\frac{\Omega_n \sigma_{1}(\mu_{2}-\mu_{1})+2\Psi_n \sigma_{2}^2}{\Omega_n^2\sigma_{1}^2+2\sigma_{2}^2}
\,,
F_n=\frac{\Omega_n \sigma_{2}(\mu_{1}-\mu_{2})+2\Psi_n \sigma_{1}^2}{\Omega_n^2\sigma_{2}^2+2\sigma_{1}^2}
\,,
\nn\\
&
G_n=
S_n \Phi_n \exp\left(-\frac{1}{2}
\left(
\left(\frac{\mu_{2}-\mu_{1}}{\sigma_{2}}\right)^2
+2\left(\frac{\Psi_n}{\Omega_n}\right)^2
-\left(
\frac{\Omega_n^2\sigma_{1}(\mu_{2}-\mu_{1})+2 \Psi_n\sigma_{2}^2}
{\Omega_n\sigma_{2}\sqrt{\Omega_n^2\sigma_{1}^2+2\sigma_{2}^2}}
\right)^2
\right)
\right)
\,,
\nn
\\
&
H_n=
T_n \Phi_n \exp\left(-\frac{1}{2}
\left(
\left(\frac{\mu_{1}-\mu_{2}}{\sigma_{1}}\right)^2
+2\left(\frac{\Psi_n}{\Omega_n}\right)^2
-\left(
\frac{\Omega_n^2\sigma_{2}(\mu_{1}-\mu_{2})+2 \Psi_n\sigma_{1}^2}
{\Omega_n\sigma_{1}\sqrt{\Omega_n^2\sigma_{2}^2+2\sigma_{1}^2}}
\right)^2
\right)
\right)
\,.
\tag{25}
\label{ecISDF_int2}
\end{align}
}
\end{tabular}
\end{table*}
\begin{table*}[!b]
\begin{tabular}{m{\textwidth}}
\hrule{
\begin{align}
&I_4
=\mu_{0}
Q\left(
\frac{\ln(\Gamma_{0})-\mu_{0}}{\sigma_{0}}\right)
+\sigma_{0}^2
\left(
\frac{1}{\sqrt{2\pi\sigma_{0}^2}}
\exp\left(
-\frac{1}{2}
\left(
\frac{\ln(\Gamma_{0})-\mu_{0}}{\sigma_{0}}\right)^2
\right)
\right)\,,
\tag{27}
\label{ecISDF1}
\end{align}
}
{
\begin{align}
&I_5
=\mu_{0}
\left(1-Q\left(
\frac{\ln(\Gamma_{0})-\mu_{0}}{\sigma_{0}}\right)\right)
+\sigma_{0}^2
\left(
\frac{-1}{\sqrt{2\pi\sigma_{0}^2}}
\exp\left(
-\frac{1}{2}
\left(
\frac{\ln(\Gamma_{0})-\mu_{0}}{\sigma_{0}}\right)^2
\right)
\right)\,,
\tag{28}
\label{ecISDF2}
\end{align}
}
{
\begin{align}
I_6
&
=
Q\left(
\frac{\ln(\Gamma_{1})-\mu_{1}}{\sigma_{1}}\right)
\left\lbrace
\mu_{2}
\left(1-Q\left(
\frac{\ln(\Gamma_{1})-\mu_{2}}{\sigma_{2}}\right)\right)
+\sigma_{2}^2
\left(
\frac{-1}{\sqrt{2\pi\sigma_{2}^2}}
\exp\left(
-\frac{1}{2}
\left(
\frac{\ln(\Gamma_{1})-\mu_{2}}{\sigma_{2}}\right)^2
\right)
\right)
\right\rbrace
\nn\\
&
+
\sum_{n=1}^{M}
G_n
\sigma_{1}
\left\lbrace
E_n
Q\left(
\frac{\frac{\ln(\Gamma_{1})-\mu_{1}}{\sigma_{1}}-E_n}{S_n}
\right)
+S_n^2
\left(
\frac{1}{\sqrt{2\pi S_n^2}}
\exp\left(
-\frac{1}{2}
\left(
\frac{\frac{\ln(\Gamma_{1})-\mu_{1}}{\sigma_{1}}-E_n}{S_n}\right)^2
\right)
\right)
\right\rbrace
\nn\\
&
+
\sum_{n=1}^{M}
G_n
\mu_{1}
Q\left(
\frac{\frac{\ln(\Gamma_{1})-\mu_{1}}{\sigma_{1}}-E_n}{S_n}
\right)
+
\sum_{n=1}^{M}
H_n
\sigma_{2}
\left\lbrace
F_n
Q\left(
\frac{\frac{\ln(\Gamma_{2})-\mu_{2}}{\sigma_{2}}-F_n}{T_n}
\right)\right.\nn\\
&\left.+T_n^2
\left(
\frac{1}{\sqrt{2\pi T_n^2}}
\exp\left(
-\frac{1}{2}
\left(
\frac{\frac{\ln(\Gamma_{2})-\mu_{2}}{\sigma_{2}}-F_n}{T_n}\right)^2
\right)
\right)
\right\rbrace
+
\sum_{n=1}^{M}
H_n
\mu_{2}
Q\left(
\frac{\frac{\ln(\Gamma_{2})-\mu_{2}}{\sigma_{2}}-F_n}{T_n}
\right)\,.
\tag{29}
\label{ecISDF3}
\end{align}
}
\end{tabular}
\end{table*}

\subsection{ISDF}
\label{subsec_isdf}
The average channel capacity for the system with the ISDF strategy can be calculated by summing up the average channel capacities of the following events:
i) the average channel capacity of the direct link when the $SD$ link SNR meets the threshold requirement, ii) the average channel 
capacity of the relayed transmission when the $SD$ link SNR fails to meet the threshold requirement and the $SR$ link SNR satisfies the threshold requirement, and 
iii) if the $SR$ link fails to achieve the SNR threshold, we assume that some rate can be achieved through the direct link even though its SNR is below $\Gamma_{0}$. {The last event ensures spectral efficiency of the ISDF system by not allowing the relay to participate if the $SR$ link has poor SNR.}  
Thus, the end-to-end average channel capacity can be expressed as 
\begin{align}
{\cal C}_{\text{ISDF}}
&=
{\mathbb E}\left[
{\cal C}(\gamma_{0}|\gamma_{0}>\Gamma_{0})\right]
+\frac{1}{2}\mbox{Pr}[\gamma_{0}<\Gamma_{0}]
{\mathbb E}\left[
{\cal C}(\gamma_{m}|\gamma_{1}>\Gamma_{1})\right]
\nn\\
&
+\mbox{Pr}[\gamma_{1}<\Gamma_{1}]
{\mathbb E}\left[{\cal C}(\gamma_{0}|\gamma_{0}<\Gamma_{0})\right]
\,.
\tag{22}
\label{ec9}
\end{align}
The PDF of $\{\gamma_m|\gamma_{1}>\Gamma_{1}\}$ needs to
be found out before evaluating the average capacity. Events $\{\gamma_m\}$ and 
$\{\gamma_{1}>\Gamma_{1}\}$ are not independent, as
$\gamma_m=\min\{\gamma_{1},\gamma_{2}\}$. 
 {Hence, the PDF of $\{\gamma_{m}|\gamma_{1}>\Gamma_{1}\}$ can be expressed as} 
\begin{align}
 &f_{\gamma_{m}|\gamma_{1}>\Gamma_{1}}(x)
 \nn\\
=& \left\{ \begin{array}{ll}
         Q\left(
         \frac{\ln\Gamma_{1}-\mu_{1}}{\sigma_{1}}
         \right)
         \frac{1}{x \sqrt{2\pi\sigma_{2}^2}}
         \exp\left(-\frac{1}{2}
         \left(
         \frac{\ln x-\mu_{2}}{\sigma_{2}}
         \right)^2 \right)
         & \mbox{ $ x \leq \Gamma_{1}$};
         \\
        Q\left(
         \frac{\ln x-\mu_{1}}{\sigma_{1}}
         \right)
         \frac{1}{ x \sqrt{2\pi\sigma_{2}^2}}
         \exp\left(-\frac{1}{2}
         \left(
         \frac{\ln x-\mu_{2}}{\sigma_{2}}
         \right)^2 \right)
\\
         +Q\left(
         \frac{\ln x-\mu_{2}}{\sigma_{2}}
         \right)
         \frac{1}{ x \sqrt{2\pi\sigma_{1}^2}}
         \exp\left(-\frac{1}{2}
         \left(
         \frac{\ln x -\mu_{1}}{\sigma_{1}}
         \right)^2 \right)
         & \mbox{$ x > \Gamma_{1}.$}
         \end{array} \right.
         \tag{23}
\label{ec12}
\end{align}


To evaluate (\ref{ec9}) in closed-form is difficult as integrals over the PDF obtained above should be solve. Hence, either the
Gaussian $Q$-function or the PDF itself needs to be approximated. 
Several approximations exist in the literature for the Gaussian $Q$-function, e.g., 
in \cite{KarLio:07, IsuRao:08}. However, these approximations are mathematically intractable in this case. 
Instead, by utilizing an approximation provided in \cite{BenCas:11} which uses only a single Gaussian function, 
we develop an approximation method with multiple Gaussian functions using non-linear least-square curve fitting technique as 
\begin{align}
Q\left(x\right)\approx
\sum_{n=1}^{N}
\Phi_n \exp\left(-\left(\frac{x-\Psi_n}{\Omega_n}\right)^2\right)\,,
\tag{24}
\label{ec13}
\end{align}
where $\Phi_n, \Psi_n,$ and $\Omega_n$ are fitting constants. Compared to the approximation provided in \cite{BenCas:11}, 
our approximation is more accurate as well as mathematically tractable.
The number of summation terms, $N$, depends on the region of interest and accuracy of the fit. 
A suitable value of $N$ and corresponding $\Phi_n, \Psi_n,$ and $\Omega_n$ are discussed later, in the results section.
 {Using this approximation and the terms defined in} (\ref{ecISDF_int2}),  {the end-to-end average channel capacity for the ISDF strategy can be evaluated as}
%
\begin{align}
{\cal C}_{\text{ISDF}}
 &
=\frac{\ln(\tau_{1}^{ \Lambda_1}\tau_{2}^{ \Lambda_2})}{\ln(2)}
\left\lbrace
Q\left(\frac{\ln(\Gamma_{0})-\mu_{0}}{\sigma_{0}}\right)
\right.
 \nn\\
 &
+\left(1-Q\left(\frac{\ln(\Gamma_{1})-\mu_{1}}{\sigma_{1}}\right)\right)
\left(1-Q\left(\frac{\ln(\Gamma_{0})-\mu_{0}}{\sigma_{0}}\right)\right)
 \nn\\
 &
+\frac{1}{2}
\left(1-Q\left(\frac{\ln(\Gamma_{0})-\mu_{0}}{\sigma_{0}}\right)\right)
\left.
\left(Q\left(\frac{\ln(\Gamma_{1})-\mu_{1}}{\sigma_{1}}\right)\right)
\right\rbrace
\nn\\
&+\frac{1}{\ln(2)}
\left\lbrace I_4
+\left(1-Q\left(\frac{\ln(\Gamma_{1})-\mu_{1}}{\sigma_{1}}\right)\right)
I_5
\right.
 \nn\\
 &
\left.
+\frac{1}{2}
\left(1-Q\left(\frac{\ln(\Gamma_{0})-\mu_{0}}{\sigma_{0}}\right)\right)
I_6
\right\rbrace\,,
\tag{26}
\label{ecISDF_CE}
\end{align}
where $I_4,I_5,$ and $I_6$ are given by (\ref{ecISDF1}), (\ref{ecISDF2}), and (\ref{ecISDF3}), respectively, shown at the bottom of the page.
\begin{table*}[!b]
\begin{tabular}{m{\textwidth}}
\hrule
{
\begin{align}
{\cal P}_{\text{ISDF}}(R_{th})
=& \left\{ \begin{array}{ll}
         Q\left(\frac{\mu_{0}-\ln\Gamma_{0}}{\sigma_{0}}\right)Q\left(\frac{\mu_{1}-\ln\Gamma_{1}}{\sigma_{1}}\right) 
         +
         Q\left(\frac{\mu_{0}-\ln\Gamma_{0}}{\sigma_{0}}\right)Q\left(\frac{\ln\Gamma_{1}-\mu_{1}}{\sigma_{1}}\right)Q\left(\frac{\mu_{2}-\ln\Gamma_{th}}{\sigma_{2}}\right)
         & \mbox{for $ \Gamma_{1} \geq \Gamma_{th}$};
         \\
         \, 
         Q\left(\frac{\mu_{0}-\ln\Gamma_{0}}{\sigma_{0}}\right)Q\left(\frac{\mu_{1}-\ln\Gamma_{1}}{\sigma_{1}}\right) 
         +Q\left(\frac{\mu_{0}-\ln\Gamma_{0}}{\sigma_{0}}\right)
         Q\left(\frac{\ln\Gamma_{th}-\mu_{1}}{\sigma_{1}}\right)Q\left(\frac{\mu_{2}-\ln\Gamma_{th}}{\sigma_{2}}\right)&\,\\
          +Q\left(\frac{\mu_{0}-\ln\Gamma_{0}}{\sigma_{0}}\right)
          \left(Q\left(\frac{\ln\Gamma_{1}-\mu_{1}}{\sigma_{1}}\right)-Q\left(\frac{\ln\Gamma_{th}-\mu_{1}}{\sigma_{1}}\right)\right)
         & \mbox{for $ \Gamma_{1} < \Gamma_{th}$}
         \end{array} \right. 
         \tag{32}
\label{e36}
\end{align}
 } 
\end{tabular}
\end{table*}

\section{Outage Probability}
\label{sec_outage}
The outage probability is defined as the probability that the instantaneous channel capacity at the destination falls below a predefined rate. In other words, when the SNR threshold requirement corresponding to the rate is not fulfilled, outage occurs.
In this section, we obtain the outage probabilities for the IDF and ISDF strategies in closed-form.

\subsection{IDF}
An outage event occurs if i) the $SD$ link fails to achieve the required rate of $R_{th}$, and ii) in the relayed transmission, the minimum of the $SR$ and the $RD$ links SNRs fails to achieve twice the required rate (as the relay is half-duplex).  
Thus, the outage probability can be mathematically obtained by multiplying the probabilities of events i) and ii) as
\begin{align}
{\cal P}_{\text{IDF}}(R_{th})\hspace*{-1.2cm}&
\nn\\
&= \mbox{Pr}\left[\gamma_{0}<\Gamma_{0}\right]
\times\Pr \left[\gamma_{m}<\Gamma_{th}\right]
\nn\\
&
=\mbox{Pr}\left[\gamma_{0}<\Gamma_{0}\right]
\left(1-\mbox{Pr}\left[\gamma_{1}>\Gamma_{th}\right]
\mbox{Pr}\left[\gamma_{2}>\Gamma_{th}\right]\right)
%
\nn\\
&
=
Q\left(
\frac{\mu_{0}-\ln\left(\Gamma_{0}\right)}
{\sigma_{0}}
\right)
\left(
1-\left( Q\left(
\frac{\ln
	\left(\Gamma_{th}\right)-\mu_{1}}
{\sigma_{1}}
\right)
\right.\right.
 \nn\\
 &
 \times
\left. \left.
Q\left(
\frac{\ln
	\left(\Gamma_{th}\right)-\mu_{2}}
{\sigma_{2}}
\right)
\right)
\right).
\tag{30}
\label{e46}
\end{align} 

\subsection{ISDF}

An outage event occurs if i) the $SD$ link fails to achieve the required rate of $R_{th}$, and ii) either the relayed transmission is overruled, or if the relayed transmission happens, it fails to achieve the required rate. Thus, 
\begin{align}
{\cal P}_{\text{ISDF}}(R_{th})
&=
{\mbox{Pr}}\left[\gamma_{0}<\Gamma_{0}\right]\nn\\
&\times\left({\mbox{Pr}}\left[\gamma_{1}<\Gamma_{1}\right]
+{\mbox{Pr}}\left[\gamma_{m}<\Gamma_{th}|\gamma_{1}\geq\Gamma_{1}\right]
\right)\,.
\tag{31}
\label{e36a}
\end{align}
Based on the values of $\Gamma_{th}$ and $\Gamma_1$, the outage probability can be expressed in closed-form as
in (\ref{e36}), placed at the bottom of this page.

It is interesting to observe that if $\Gamma_{1}$ is set to $\Gamma_{th}$, outage probabilities for the IDF and ISDF strategies in (\ref{e46}) and (\ref{e36}), respectively, become the same.  

 \begin{table*}[!b]
\begin{tabular}{m{\textwidth}}
\hrule
{
\begin{align}
BER_{{\text{IDF}}}
&=
\sum_{n=1}^{M}
\sum_{j=1}^{2}
\frac{2\, \Lambda_j\,\alpha_n}{\sigma_{0}\sqrt{2}A_{n,{1}}}
\exp\left(
-\left(
C_{n,j,{1}}-\left(\frac{B_{n,j,{1}}}{A_{n,1}}\right)^2
\right)
\right)
\left[
Q\left(\sqrt{2}\left(A_{n,1}\ln(\sqrt{\tau_{j}\Gamma_{0}})-\frac{B_{n,j,{1}}}{A_{n,{1}}}\right)\right)
\right]
\nn\\
&
+
\left(
1-Q\left(
\frac{\ln\Gamma_{0}-\mu_{0}}{\sigma_{0}}
\right)
\right)
\left\lbrace
\Bigg(
1-
\sum_{n=1}^{M}
\sum_{j=1}^{2}
\frac{2\, \Lambda_j\,\alpha_n}{\sigma_{1}\sqrt{2}A_{n,{2}}}
\right.
\left.
\exp\left(
-\left(
C_{n,j,{2}}-\left(\frac{B_{n,j,{2}}}{A_{n,2}}\right)^2
\right)
\right)
\right.
\Bigg)
\nn\\
&
\times
\Bigg(
\sum_{n=1}^{M}
\sum_{j=1}^{2}
\frac{2\, \Lambda_j\,\alpha_n}{\sigma_{2}\sqrt{2}A_{n,{3}}}
\left.
\exp\left(
-\left(
C_{n,j,{3}}-\left(\frac{B_{n,j,{3}}}{A_{n,3}}\right)^2
\right)
\right)
\right.
\Bigg)
+
\Bigg(
\sum_{n=1}^{M}
\sum_{j=1}^{2}
\frac{2\, \Lambda_j\,\alpha_n}{\sigma_{1}\sqrt{2}A_{n,{2}}}
\nn
\\
&
\times \left.
\exp\left(
-\left(
C_{n,j,{2}}-\left(\frac{B_{n,j,{2}}}{A_{n,2}}\right)^2
\right)
\right)
\right.
\Bigg)
\Bigg(
1-
\sum_{n=1}^{M}
\sum_{j=1}^{2}
\frac{2\, \Lambda_j\,\alpha_n}{\sigma_{2}\sqrt{2}A_{n,{3}}}
\nn
\left.
\exp\left(
-\left(
C_{n,j,{3}}-\left(\frac{B_{n,j,{3}}}{A_{n,3}}\right)^2
\right)
\right)
\right.
\Bigg)\Bigg\}. \nn
\tag{39}
\label{e39}
\end{align} 
}
\end{tabular}
\end{table*}

\begin{table*}[!b]
\begin{tabular}{m{\textwidth}}
\hrule
{
\begin{align}
BER_{{\text{ISDF}}}
&=
\sum_{n=1}^{M}
\sum_{j=1}^{2}
\frac{2\, \Lambda_j\,\alpha_n}{\sigma_{0}\sqrt{2}A_{n,{1}}}
\exp\left(
-\left(
C_{n,j,{1}}-\left(\frac{B_{n,j,{1}}}{A_{n,1}}\right)^2
\right)
\right)
\left[
Q\left(\sqrt{2}\left(A_{n,1}\ln(\sqrt{\tau_{j}\Gamma_{0}})-\frac{B_{n,j,{1}}}{A_{n,{1}}}\right)\right)
\right]
\nn\\
&
+
\left(
1-Q\left(
\frac{\ln\Gamma_{1}-\mu_{1}}{\sigma_{1}}
\right)
\right)
\sum_{n=1}^{M}
\sum_{j=1}^{2}
\frac{2\, \Lambda_j\,\alpha_n}{\sigma_{0}\sqrt{2}A_{n,{1}}}
\exp\left(
-\left(
C_{n,j,{1}}-\left(\frac{B_{n,j,{1}}}{A_{n,1}}\right)^2
\right)
\right)
 \nn\\
&
\times
\left[1-
Q\left(\sqrt{2}\left(A_{n,1}\ln(\sqrt{\tau_{j}\Gamma_{0}})-\frac{B_{n,j,{1}}}{A_{n,{1}}}\right)\right)
\right]
+
\left(
1-Q\left(
\frac{\ln\Gamma_{0}-\mu_{0}}{\sigma_{0}}
\right)
\right)
\nn\\
&
\times
\left\lbrace
\Bigg(
1-
\sum_{n=1}^{M}
\sum_{j=1}^{2}
\frac{2\, \Lambda_j\,\alpha_n}{\sigma_{1}\sqrt{2}A_{n,{2}}}
\right.
\left.
\exp\left(
-\left(
C_{n,j,{2}}-\left(\frac{B_{n,j,{2}}}{A_{n,2}}\right)^2
\right)
\right)
\right.
\Bigg)
\Bigg(
\sum_{n=1}^{M}
\sum_{j=1}^{2}
\frac{2\, \Lambda_j\,\alpha_n}{\sigma_{2}\sqrt{2}A_{n,{3}}}
\nn
\\
&
\times\left.
\exp\left(
-\left(
C_{n,j,{3}}-\left(\frac{B_{n,j,{3}}}{A_{n,3}}\right)^2
\right)
\right)
\right.
\left(
Q\left(
\frac{\ln\Gamma_{1}-\mu_{1}}{\sigma_{1}}
\right)
\right)\Bigg)\nn
+
\Bigg(
\sum_{n=1}^{M}
\sum_{j=1}^{2}
\frac{2\, \Lambda_j\,\alpha_n}{\sigma_{1}\sqrt{2}A_{n,{2}}}
\nn
\\
&
\times
\left.
\exp\left(
-\left(
C_{n,j,{2}}-\left(\frac{B_{n,j,{2}}}{A_{n,2}}\right)^2
\right)
\right)
\right.
\Bigg)
\Bigg(
1-
\sum_{n=1}^{M}
\sum_{j=1}^{2}
\frac{2\, \Lambda_j\,\alpha_n}{\sigma_{2}\sqrt{2}A_{n,{3}}}
\left.
\exp\left(
-\left(
C_{n,j,{3}}-\left(\frac{B_{n,j,{3}}}{A_{n,3}}\right)^2
\right)
\right)
\right.
\left(
Q\left(
\frac{\ln\Gamma_{1}-\mu_{1}}{\sigma_{1}}
\right)
\right)
\Bigg)\Bigg\}.
\tag{41}
\label{e35a}
\end{align}
}
\end{tabular}
\end{table*}

\section{Average Bit Error Rate}
\label{sec_ABER}
In this section, we derive the average BER expressions for the relaying schemes assuming binary phase-shift keying (BPSK) signaling. 

\subsection{IDF}
According to the IDF strategy under consideration, a bit error can occur either in the direct or in the relayed transmission to $D$.
A bit error in the direct transmission can occur even if its SNR exceeds the required threshold. 
Now, a bit error in the relayed transmission can occur only if one of the links between the $SR$ or the $RD$ is in error when the $SD$ link SNR does not meet the required threshold.
Thus, the average BER for binary signaling can therefore be obtained by summing up the probabilities of all the above events as
\begin{align}
BER_{{\text{IDF}}}
&=
{\mathbb E}\left[P_e(\gamma_{0}|\gamma_{0}\geq\Gamma_{0})\right]
+
\mbox{Pr}\left[\gamma_{0}<\Gamma_{0}\right]
 \nn\\
&\times
\left(
\left(1-{\mathbb E}\left[P_e(\gamma_{1})\right]\right)
{\mathbb E}\left[P_e(\gamma_{2})\right]
\right.\nn\\
&+
\left.
{\mathbb E}\left[P_e(\gamma_{1})\right]
\left(1-{\mathbb E}\left[P_e(\gamma_{2})\right]\right)
\right).
\tag{33}
\label{e38}
\end{align}
In the above expression, we have discarded the event of two consecutive error detection in relayed transmission as this would result in a correct decision at the destination for BPSK. 

For equiprobable BPSK signaling scheme,
$P_e(v)$ is given by \cite{DuMaSc:15}
\begin{align}
P_e(x)=\sum_{j=1}^{2} \Lambda_jQ\left(\sqrt{\tau_j\,x} \right).
\tag{34}
\label{e28}
\end{align}
Thus, to get a closed-form solution for (\ref{e38}),
we need the solution of the following type of integral
\begin{align}
{\mathbb E}[P_e(x|x_1<X\leq x_2)]
\hspace{-3cm}
&
\nn\\
&=
\sum_{j=1}^{2}\int_{x_1}^{x_2}
 \Lambda_j\,Q\left(\sqrt{\tau_{j} x}\right)
 \nn\\
 &\times
\frac{1}{x\sqrt{2\pi\sigma_{i}^2}}
\exp{\left(-\frac{1}{2}\left(
\frac{\ln{(x)}-\mu_{i}}{\sigma_{i}}\right)^2
\right)}{\mbox{d}}x
&
\nn\\
&=
\sum_{j=1}^{2}\int_{\ln(\sqrt{\tau_{j}x_1})}^{\ln(\sqrt{\tau_{j}x_2})}
 \Lambda_j\,Q\left(\exp(t)\right)
 \nn\\
 &\times
\frac{2}{\sqrt{2\pi\sigma_{i}^2}}
\exp{\left(-\frac{1}{2}\left(
\frac{2t-\ln(\tau_j)-\mu_{i}}{\sigma_{i}}\right)^2
\right)}{\mbox{d}}t.
\tag{35}
\label{e32}
\end{align}
It is difficult to evaluate the above integral in closed-form, as it 
contains a function of the form $Q(\exp(x))$. 
Therefore, using the same curve fitting technique proposed in the Section \ref{subsec_isdf}, 
the approximation of the function $Q(\exp(x))$ is given by 
\begin{align}
Q(\exp(x))\approx\sum_{n=1}^{M}
\alpha_n \exp\left(-\left(\frac{x-\beta_n}{\delta_n}\right)^2\right)\,,
\tag{36}
\label{e33}
\end{align}
where $\alpha_n, \beta_n,$ and $\delta_n$ are fitting constants. 
The number of summation terms, $M$, depends on the region of interest and accuracy of the fit.
Using this approximation, the integral in (\ref{e32}) can be evaluated in closed-form
as 
\begin{align}
&{\mathbb E}[P_e(x|x_1<X\leq x_2)]\nn\\
&\approx
\sum_{n=1}^{M}
\sum_{j=1}^{2}
\frac{2\, \Lambda_j\,\alpha_n}{\sigma_{i}\sqrt{2}A_{n,i}}
\exp\left(
-\left(
C_{n,j,i}-\left(\frac{B_{n,j,i}}{A_{n,i}}\right)^2
\right)
\right)
 \nn\\
 &
 \times
\left\{
Q\left(\sqrt{2}\left(A_{n,i}\ln(\sqrt{\tau_{j}x_1})-\frac{B_{n,j,i}}{A_{n,i}}\right)\right)
\right.
\nn\\
&
\left.
-
Q\left(\sqrt{2}\left(A_{n,i}\ln(\sqrt{\tau_{j}x_2})-\frac{B_{n,j,i}}{A_{n,i}}\right)\right)
\right\}\,,
\tag{37}
\label{e34}
\end{align}
where
\begin{align}
A_{n,i}&=\sqrt{\frac{1}{\delta_n^2}+
\frac{2}{\sigma_{i}^2}}
\,,
\,\,
B_{n,j,i}
=\frac{\beta_n}{\delta_n^2}+
\frac{\ln(\tau_j)+\mu_{i}}{\sigma_{i}^2}
\,,
\nn\\
\,\,
&C_{n,j,i}
=\frac{\beta_n^2}{\delta_n^2}+
\frac{\left(\ln(\tau_j)+\mu_{i}\right)^2}{2\sigma_{i}^2}
\,.
\tag{38}
\label{e35}
\end{align}
Finally, using (\ref{e23}) and (\ref{e34}),
the average BER in (\ref{e38}) can be expressed in approximate closed-form as
 in (\ref{e39}) at the bottom of the next page.

\subsection{ISDF}   
The average BER for the system with ISDF strategy can be calculated by summing up the average BER of the following events:
i) the average BER of the direct link when the $SD$ link SNR meets the threshold requirement, ii) the average BER of the relayed transmission when the $SD$ link SNR fails to fulfill the threshold requirement and the $SR$ link SNR achieves the threshold requirement, and 
iii) if the $SR$ link fails to achieve the SNR threshold, we assume that decision can be taken based on the symbol received through the direct link even though its SNR is below $\Gamma_{0}$. 
Thus, the average BER for binary signaling can be obtained by summing up the probabilities of all the above events as 
\begin{align}
BER_{{\text{ISDF}}}
\hspace*{-1.5cm}
&
\nn\\
&=
{\mathbb E}\left[P_e(\gamma_{0}|\gamma_{0}\geq\Gamma_{0})\right]
 \nn\\
&
+
\mbox{Pr}\left[\gamma_{0}<\Gamma_{0}\right]\Big(
\left(1-{\mathbb E}\left[P_e(\gamma_{1}|\gamma_{1}\geq\Gamma_{1})\right]\right)
{\mathbb E}\left[P_e(\gamma_{2}|\gamma_{1}\geq\Gamma_{1})\right]
\nn\\
&
+
{\mathbb E}\left[P_e(\gamma_{1}|\gamma_{1}\geq\Gamma_{1})\right]
\left(1-{\mathbb E}\left[P_e(\gamma_{2}|\gamma_{1}\geq\Gamma_{1})\right]\right)
\Big)\nn\\
&
+\mbox{Pr}\left[\gamma_{1}<\Gamma_{1}\right]
{\mathbb E}\left[P_e(\gamma_{0}|\gamma_{0}<\Gamma_{0})\right]
.
\tag{40}
\label{e27}
\end{align}
Using (\ref{e23}) and (\ref{e34}),
the average BER in (\ref{e27}) can be expressed in approximate closed-form as 
in (\ref{e35a}) at the bottom of the page.

%

\section{Relay Usage}
\label{relay_usgae}
The more the relay is used for data transmission, the poorer the spectral efficiency, and the more the additional complexity and delay required in data processing. {However, the more the relay is used, the better the outage performance.}  
Hence, in this section, {\em the fraction of times the relay is in use} is calculated for both IDF and ISDF strategies.

In the IDF, the relay is used only if the $SD$ link SNR fails to attain the threshold requirement. Thus, the fraction of times the relay is in use can be obtained simply by finding the probability that the $SD$ link SNR fails to attain the threshold requirement. In the ISDF, the relay is used if the $SD$ link SNR fails to attain the threshold requirement while the $SR$ link SNR attains it. Hence, the fraction of times the relay is in use for IDF and ISDF can be expressed as
\begin{align}
N_{\text{IDF}}&=
\mbox{Pr}[\gamma_{0}<\Gamma_{0}]
=1-Q\left(
\frac{\ln\left(\Gamma_{0}\right)-\mu_{0}}
{\sigma_{0}}
\right)\,,
\tag{42}
\label{eq_avguse1}
\nn\\
N_{\text{ISDF}}
&
=
\mbox{Pr}[\gamma_{0}<\Gamma_{0}]\mbox{Pr}[\gamma_{1}>\Gamma_{1}]
 \nn\\
 &
\hspace*{-0.5cm}
= \left(1-Q\left(
\frac{\ln\left(\Gamma_{0}\right)-\mu_{0}}
{\sigma_{0}}
\right)\right)
Q\left(
\frac{\ln\left(\Gamma_{1}\right)-\mu_{1}}
{\sigma_{1}}
\right)\,,
\tag{43}
\label{eq_avguse2}
\end{align}
respectively.

\section{Optimum Power Allocation}
\label{sec_power}
Instead of allocating equal power to $S$ and $R$, the power can be allocated judiciously to improve the system performance. In this section, we investigate the power allocation problem to minimize the outage probability. 
Towards this goal, we first check whether or not the outage probability is a convex function of the power allocation factor, $p_f$. 
Then, we find the global optimum power allocation factor \cite{book_boyd}. 
{One can observe from} Section \ref{sec_outage} that the outage probability is the same for both IDF and ISDF when $\Gamma_{1}=\Gamma_{th}$. Hence, power allocation is solved only for this case. 

\begin{table}[t]
	\caption{Parameters for (\ref{ec13}) and (\ref{e33}) from curve fitting for $M=N=7$.}
	\begin{center}
\scalebox{0.78}{
		\begin{tabular}{| c | c | c|c| c | c|c|}
			\hline $n$ & $\alpha_{n}$ & $\beta_{n}$ & $\delta_{n}$ & $\Phi_{n}$ & $\Psi_{n}$ & $\Omega_{n}$\\
			\hline
			$1$ & 0.4665 & -5.37 & 2.174 & 0.9302 & -5.48 & 2.833\\
			\hline
			$2$ & -0.0007029 & -3.674 & 0.1178 & 0.0001404 & -1.157 & 0.01036\\
			\hline
			$3$ & 0.0165 & -3.141 & 0.0004957 & 0.0007985  & -1.381 & 0.021 \\
			\hline
			$4$ & 0.2831 & -2.998 & 1.458 & -0.001064 & -0.9854 & 0.158 \\
			\hline
			$5$ & 0.2113 & -1.764 & 1.06 & 0.00196 & -1.699 & 0.173\\
			\hline
			$6$ & 0.1742 & -0.8425 & 0.837 & 0.4171 & -0.7018 & 1.535 \\
			\hline
			$7$ & 0.07986 & -0.1109 & 0.6399 & 0.5843 & -2.347 & 1.96\\
			\hline
		\end{tabular}
		}
	\end{center}
	\label{tab1}
\end{table}
 
The outage probability in (\ref{e46}) can be expressed as a function of the power allocation factor, $p_f$, as

\begin{align}
\hspace*{-0.5cm}
{\cal P}_{IDF}(p_f)
&
=
Q\left(v_{0}\ln(p_f)-u_{0}\right)
\left[
Q\left(v_{1}\ln(p_f)-u_{1}\right)
\right.
\nn\\
&\left.
+Q\left(v_{2}\ln(1-p_f)-u_{2}\right)
\right]
- Q\left(v_{0}\ln(p_f)-u_{0}\right)
\nn\\
&\times 
Q\left(v_{1}\ln(p_f)-u_{1}\right)
Q\left(v_{2}\ln(1-p_f)-u_{2}\right)\,,
\tag{44}
\label{ep2}
\end{align}
where
\begin{align}
&v_{i}=\frac{1}{\sigma_{i}},
u_{i}=\frac{\ln(\Gamma_{i})-2 \Xi_{i}-\ln(P_T/(N_0 P_{L}d_{i}))}{\sigma_{i}}
.
\tag{45}
\label{ep1}
\end{align}
As the $Q$-function values are always less than unity, the product of three $Q$-functions will be very small 
when compared to the other terms in (\ref{ep2}). Hence, we get an approximation by 
neglecting this product, as 
\begin{align}
{\cal P}_{\text{IDF}}(p_f)&=
Q\left(v_{0}\ln(p_f)-u_{0}\right)
\left[
Q\left(v_{1}\ln(p_f)-u_{1}\right)
\right.
\nn\\
&\left.
+Q\left(v_{2}\ln(1-p_f)-u_{2}\right)
\right].
\tag{46}
\label{ep7}
\end{align}
The approximation is more accurate when the argument of the $Q$-function is positive and large at higher $P_T$. 
This approximation will help up to show the convexity of the power allocation problem easily, as opposed to the actual problem.


 {The first and second-order derivative of} (\ref{ep7}),  {with respect to $p_f$, are obtained as}
\begin{align}
{\cal P}_{\text{IDF}}^{'}(p_f)
&=
Q^{'}\left(v_{0}\ln(p_f)-u_{0}\right)
\left[
Q\left(v_{1}\ln(p_f)-u_{1}\right)
\right.
\nn\\
&\left.
+Q\left(v_{2}\ln(1-p_f)-u_{2}\right)
\right]
\nn\\
&+
Q\left(v_{0}\ln(p_f)-u_{0}\right)
\left[
Q^{'}\left(v_{1}\ln(p_f)-u_{1}\right)
\right.
\nn\\
&\left.
+Q^{'}\left(v_{2}\ln(1-p_f)-u_{2}\right)
\right],
\tag{47}
\label{ep8}
\\
{\cal P}_{\text{IDF}}^{''}(p_f)
&=
Q^{''}\left(v_{0}\ln(p_f)-u_{0}\right)
\left[
Q\left(v_{1}\ln(p_f)-u_{1}\right)
\right.
\nn\\
&\left.
+Q\left(v_{2}\ln(1-p_f)-u_{2}\right)
\right]
\nn\\
&
+
2Q^{'}\left(v_{0}\ln(p_f)-u_{0}\right)
\left[
Q^{'}\left(v_{1}\ln(p_f)-u_{1}\right)
\right.
\nn\\
&\left.
+Q^{'}\left(v_{2}\ln(1-p_f)-u_{2}\right)
\right]
\nn\\
&+
2Q\left(v_{0}\ln(p_f)-u_{0}\right)
\left[
Q^{''}\left(v_{1}\ln(p_f)-u_{1}\right)
\right.
\nn\\
&\left.
+Q^{''}\left(v_{2}\ln(1-p_f)-u_{2}\right)
\right],
\tag{48}
\label{ep9}
\end{align}
 {where $Q^{'}(\cdot)$ and $Q^{''}(\cdot)$ denote the first and second-order derivative of the $Q$-function}  
\cite{Gu:63}. 
%
It is still difficult to confirm straightway that the approximate ${\cal P}_{\text{IDF}}^{''}(p_f)$  is positive $\forall p_f$. 
Instead, we can show from (\ref{ep9}) that when $P_T$ is sufficiently high, ${\cal P}_{\text{IDF}}^{''}(p_f)$ is positive.
We can see from (\ref{ep1}) that the arguments of $Q$-functions are positive, i.e.,  
$\left(v_{i}\ln(p_f)-u_{i}\right)>0$ and $\left(v_{i}\ln(1-p_f)-u_{i}\right)>0$. 
Thus, (\ref{ep9}) can be easily shown to always be positive if $P_T$ is sufficiently high so that the 
received SNR is greater than some quantity, i.e., 
\begin{align}
&P_T/(N_0 P_{L}d_{2})>
\exp\left(\max\{\theta_i, \phi_i\}\right),
\nn
\end{align}
where $\theta_i=\ln(\Gamma_{i})-2\Xi_i-v_i \ln(p_f)$ and $\phi_i=\ln(\Gamma_{i})-2\Xi_i-v_i \ln(1-p_f)$.

At this point we have shown analytically that at sufficiently high SNR, the approximated ${\cal P}_{\text{IDF}}(p_f)$ 
is a convex function of $p_f$. Therefore, a global minimum for ${\cal P}_{\text{IDF}}(p_f)$ exists. 
By equating ${\cal P}_{\text{IDF}}^{'}(p_f)$ to zero, $p_f$ can be evaluated.
It is, however, difficult to evaluate $p_f$ in closed-form due to the complexity of the equation involved. Hence, we compute it numerically.  

\section{Results and Discussions}
\label{sec_results}

Numerical and simulation results are presented in this section to validate the analysis. 
The distance between $S$ and $D$ is assumed to be between $200$ and $400$ meters.
This range of distances best suits small PLC system environment, such as home automation and load control applications.
Depending on the power distribution network (number of paths in multipath model), in general, $\xi_i$ lies in between $2$ dB to $5$ dB \cite{GuCeAr:11}. 
High value of $\xi_i$ indicates high fluctuation in the received signal power \cite{PaCaKaTh:03,GuCeAr:11}.  
Here, $\xi_i$ is represented in dB, for all $i$, where the conversion from natural scale to dB scale is given by $\xi_i (\text{dB}) = 10\xi_i/ \ln{10}$.  
It has been shown that the path-loss in PLC ranges from $40$ to $80$ dB/km \cite{Ho:98}; hence, $P_L=50$ dB/km is chosen for the analysis. 
Unless otherwise specified, the value of the impulsive noise parameters are $ \Lambda=0.1$ and $\eta=10$. 
For the approximation presented in (\ref{ec13}), $N=7$ is found to be sufficient with root mean squared error (RMSE) of $7.264\times10^{-4}$ and sum of squares due to error (SSE) of $5.171\times10^{-4}$.
Similarly, for (\ref{e33}), $M=7$ is sufficient with RMSE of $6.931\times10^{-4}$ 
and SSE of $4.708\times10^{-4}$ for the curve fitting. 
The fitting constants corresponding to (\ref{ec13}) and (\ref{e33}) are given in Table \ref{tab1}. 
{In this section, results in Figs. 3-8 are obtained using MATLAB, by averaging over 1 million channel realizations.}
\begin{figure}
	 \begin{center}	\psfig{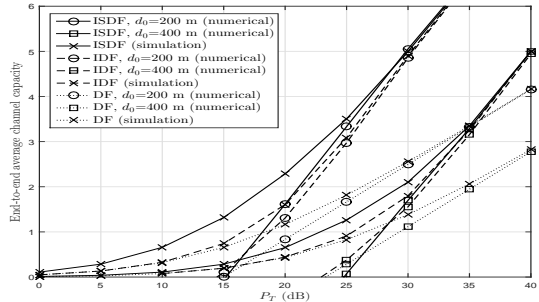}
	\caption{End-to-end average channel capacity versus total transmit power with $\xi_i =3$ dB,  
	${P}_{L}=50$ dB/km, $p_f=0.5$, $d_f=0.1$, $ \Lambda = 0.1$, $\eta = 10$, $R_{th}=4$ bits/sec/Hz, and $\Gamma_{1}=\Gamma_{th}$ for different $d_{0}$.}
	\label{fig_c}
\end{center}
\end{figure}

\begin{figure}
\begin{center}
\psfig{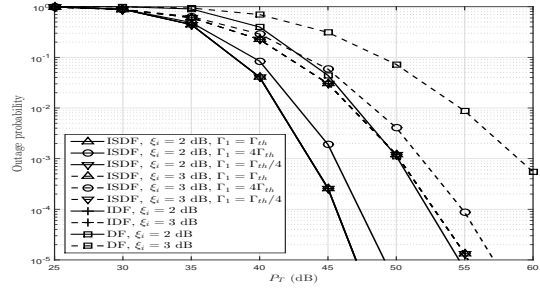}
\caption{ {Outage probability versus total transmit power with $d_{0}=400$ m, $P_L=50$ dB/km, $ \Lambda = 0.1$,  $\eta = 10$, $R_{th}=4$ bits/sec/Hz, $p_f=0.5$, and $d_f=0.5$ for various $\xi_i$ and $\Gamma_1$.}}
\label{fig_5A}
\end{center}
\end{figure}

Fig. \ref{fig_c} compares the end-to-end average channel capacity of the proposed strategies with the traditional DF strategy.
The numerical results for IDF and ISDF are obtained using the approximate closed-form expressions derived in 
(\ref{ec8}) and (\ref{ecISDF_CE}), respectively. For traditional DF, it is obtained from \cite{DuMaSc:15}.
Numerical results are in agreement with simulation results at high SNR, thus validating our high SNR approximate analysis.
In general, {Fig.} \ref{fig_c} {shows} that for a given $P_T$, the average channel 
capacity decreases with the increase in $d_{0}$ due to the path loss. 

{It can be seen} that, at relatively low $P_T$, the ISDF has the highest average channel capacity and DF has the lowest, at a given $d_{0}$. 
Though both IDF and ISDF are incremental relaying techniques, as there is an extra check on the SNR at $R$ 
for ISDF, its capacity is greater than that of IDF. On the other hand, average channel capacity of the traditional 
DF relaying is the lowest, as it is not 
an incremental relaying technique. Furthermore, average capacity performances become the same for all strategies at relatively high $P_T$ as they all follow direct transmission. Moreover, by checking the average capacity at $d_{0}=200$ m, {we observe} that ISDF achieves  $R_{th}=4$ bits/sec/Hz at a lower $P_T$ than IDF. However, at $d_{0}=400$ m, both require the same $P_T$. 
Thus, it can be concluded that ISDF outperforms IDF at comparatively low $P_T$ and $d_{0}$.


Fig. \ref{fig_5A} shows the outage probability versus ${P}_{T}$ for the proposed strategies by changing $\Gamma_1$ for a given $\xi_i$. {The general observations illustrate} that to achieve a desired outage performance, the total transmit power requirement increases with the increase in $\xi_i$. For example, for $\Gamma_{1}=\Gamma_{th}$, to achieve an outage probability of $10^{-3}$, the ISDF requires approximately $7$ dB more SNR at $\xi_i=3$ as compared to $\xi_i=2$. This is reasonable because as the channel quality degrades, the transmit power requirement increases to achieve the same performance. 
It is also seen that when $\Gamma_1$ is less than $\Gamma_{th}$, the outage performance remains the same with the increase in $\Gamma_1$. However, 
the outage performance degrades with increasing $\Gamma_{1}$ beyond $\Gamma_{th}$. 
As $\Gamma_{1}$ increases while remaining below $\Gamma_{th}$, 
the relay usage decreases. However, it does not change the outage performance because 
all relayed transmissions with $\Gamma_{1}\leq\gamma_1<\Gamma_{th}$ will be in outage as per 
the definition in Section \ref{sec_outage}. 
On the other hand, if $\Gamma_{1}$ is beyond $\Gamma_{th}$, the relay will force the system 
to go into outage by blocking the transmissions with $\Gamma_{th}<\gamma_1\leq\Gamma_{1}$.  
When $\Gamma_{1}=\Gamma_{th}$, both the IDF and the ISDF have identical outage performance.  
Thus, it can be concluded that ISDF behaves the same as IDF when $\Gamma_{1} \leq \Gamma_{th}$.

\begin{figure}
\begin{center}
	\psfig{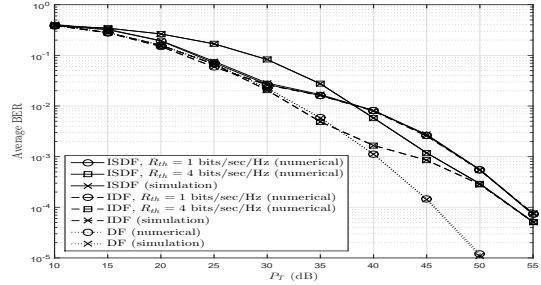}
	\caption{Average BER versus total transmit power with
		$\xi_i =3$ dB, $d_{0}=400$ m, ${P}_{L}=50$ dB/km, $p_f=0.5$, $d_f=0.5$, $ \Lambda = 0.1$, $\eta = 10$, and $\Gamma_{1}=\Gamma_{th}$ 
		for different $R_{th}$.}
	\label{fig_3}
\end{center}
\end{figure}

Fig. \ref{fig_3} shows the average BER versus ${P}_{T}$ 
for the proposed strategies 
and the traditional DF strategy for different $R_{th}$.
The numerical results for the IDF and ISDF strategies are obtained using the approximate closed-form expressions derived in (\ref{e39}) and (\ref{e35a}),
 respectively, 
whereas results for the traditional dual-hop DF scheme are obtained from \cite{DuMaSc:15}.
Numerical results are in agreement with simulation results, thus validating our analysis. 
In general, {Fig.} \ref{fig_3} {shows} that the average BER improves with the increase in $P_T$.
Furthermore, for the proposed strategies, it can be seen that when
$P_T$ is low, the average BER degrades with the increase in $R_{th}$;
however, when $P_T$ is high, it improves with the increase in $R_{th}$. 
Moreover, at very high $P_T$, the BER curves merge for different $R_{th}$. 
This is because the system tends to follow the direct transmission at such high $P_T$ values. 
This is an interesting observation as with the increase in $R_{th}$, intuitively, the average BER should degrade at
all SNRs. When $R_{th}$ increases, the average BER decreases at a lower $P_T$ as neither the $SD$ nor the $SR$ can overcome the increased
SNR threshold at $D$ and $R$, respectively. If $P_T$ is increased further, $R$ can eventually overcome the required SNR threshold
due to comparatively low path loss and increased received power at it, hence, this observation. 
Furthermore, it can be observed that the traditional DF is superior over the proposed strategies. The traditional DF, however, has very low spectral efficiency 
(to be discussed later in detail).

\begin{figure}
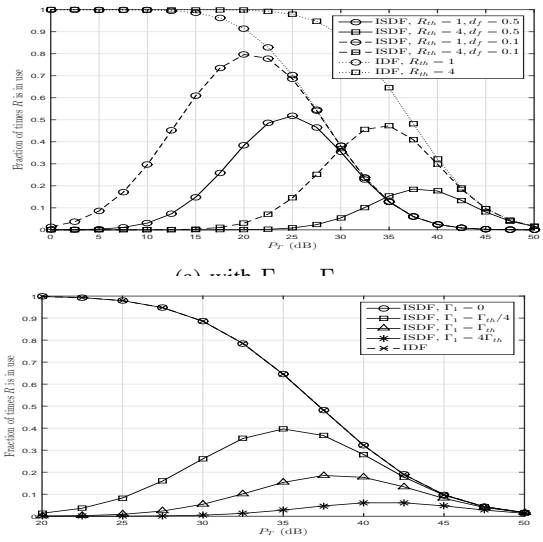

\begin{center}
\begin{subfigure}[b]{0.5\textwidth}
\hspace*{.65 cm}
\psfig{file=fig6a_ISJ-RE-18-06780.eps,width=7cm,height=3.25cm}
\caption{with $\Gamma_{1}=\Gamma_{th}$}
\label{fig_4}
\end{subfigure}
\begin{subfigure}[b]{0.5\textwidth}
\hspace*{.66 cm}\psfig{file=fig6b_ISJ-RE-18-06780.eps,width=7cm,height=3.25cm}
\caption{with $d_f=0.5$ and $R_{th}=4$ bits/sec/Hz}
\label{fig_4A}
\end{subfigure}
\caption{Fraction of times $R$ is in use versus $P_T$ with
		$\xi_i =3$ dB, $d_{0}=400$ m, ${P}_{L}=50$ dB/km, $p_f=0.5$, $ \Lambda = 0.1$, and $\eta = 10$ 
		for (a) different $R_{th}$ and $d_f$ values and (b) different $\Gamma_{1}$ values.}
\end{center}
\end{figure}

In Figs. \ref{fig_4} and \ref{fig_4A}, the fraction of times $R$ is in use for transmission versus $P_T$ is plotted for the proposed 
strategies using (\ref{eq_avguse1}) and (\ref{eq_avguse2}), respectively, by varying $R_{th}$, $d_f$, and $\Gamma_{1}$. 
{Observations reveal} that with the increase in $P_T$, the curves for IDF are approaching zero from unity. 
On the other hand, the curves for ISDF are bell-shaped and never reach unity. 
Initially, $R$ is used every time when the $SD$ link fails to achieve $R_{th}$ at low SNR in IDF. 
Later, with the increase in $P_T$, the relayed transmission ceases, and hence the observation.
However, due to threshold-selection, the relay usage never achieves unity in ISDF.  As $P_T$ increases, 
the relay usage initially increases due to improved $SR$ link quality, and then decreases due 
to better direct link quality, and hence, the bell-shaped curves for ISDF.

Next, {Fig.} \ref{fig_4} {illustrates} for the ISDF strategy that as $d_f$ increases at a given $R_{th}$, 
the corresponding curves shift towards the right and its maximum also reduces. 
As the length of the $SR$ link increases, the received SNR at $R$ decreases, which in turn reduces $N_{\text{ISDF}}$.
On the other hand, in the IDF strategy, $R$ is used every time irrespective of its placement as there is no threshold-selection.

Furthermore, as $N_{\text{ISDF}}$ is always less than or equal to $N_{\text{IDF}}$, 
the average channel capacity of the ISDF is always more than or equal to that of IDF in Fig. \ref{fig_c}. 
Thus, it can be concluded that ISDF is spectrally more efficient than both the IDF and the traditional DF relaying. 

Moreover, we can observe that in the case of ISDF strategy at a given $d_f$ and beyond a certain $P_T$, 
$N_{\text{ISDF}}$ for $R_{th}=4$ becomes more than that for $R_{th}=1$ due to the bell-shape. 
This justifies the crossovers of the average BER plots for the same $d_f$ in Fig. \ref{fig_3}. 
Thus, although the spectral efficiency decreases at higher $P_T$ when $R_{th}$ increases, interestingly, the average BER improves.
On the contrary, in case of the IDF strategy, $N_{\text{IDF}}$  for $R_{th}=1$ is always less than that for $R_{th}=4$. 
This justifies the nature of the BER curves in Fig. \ref{fig_3} for IDF.
 
{Fig.} \ref{fig_4A} {shows} that as $\Gamma_1$ increases, the average relay usage decreases. Moreover, by comparing the observations in Figs. \ref{fig_5A} and \ref{fig_4A}, it can be concluded that increasing the value of $\Gamma_{1}$ while keeping it below $\Gamma_{th}$ improves the spectral efficiency without compromising the outage performance. However, increasing its value beyond $\Gamma_{th}$ degrades the outage performance. Thus, it can be concluded that using ISDF with $\Gamma_{1}=\Gamma_{th}$ maximizes the spectral efficiency without compromising the outage performance.  

\begin{figure}
	\begin{center}
	\hspace*{.5 cm}
	\psfig{file=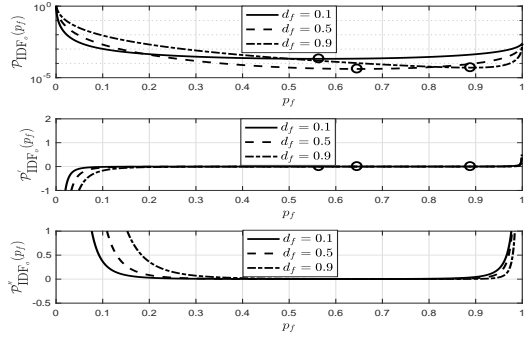,width=8cm,height=5cm}
	\caption{Outage probability, first-order and second-order derivative of the outage probability 
		versus $p_f$ for different $d_{1}$ with $P_T=50$ dB, 
		$\xi_i =4$ dB $\forall i$, $d_{0}=200$ m, ${P}_{L}=80$ dB/km, $ \Lambda = 0.1$, $\eta = 10$, $R_{th}=2$, and $\Gamma_{1}=\Gamma_{th}$.}
	\label{FIG_POW1}
\end{center}
\end{figure}

In Fig. \ref{FIG_POW1}, the outage probability and its first-order derivative (${{\cal P}^{'}_{\text{IDF}}}(p_f)$) and second-order derivative (${{\cal P}^{''}_{\text{IDF}}}(p_f)$) are shown
as a function of power allocation factors for different 
preassigned distances $d_{1}$ and $d_{2}$. Clearly, the figure shows that the second-order derivative of the outage probability is always positive, confirming the convexity of the outage probability.
The points at which the first-order derivative becomes zero, which is shown by circles,  is the point at which minimum is achieved in ${\cal P}_{\text{IDF}}(p_f)$. 
Interestingly, we can observe that the minimum point on ${\cal P}_{\text{IDF}}(p_f)$ shifts towards right as $d_f$ 
increases from 0.1 to 0.9. This means that as the distance between $S$ and $R$ increases, 
more power should be allocated to $S$. This is reasonable since as $d_{1}$ increases, 
the path loss degrades the $SR$ link, and hence, more power should be assigned to $S$ to 
compensate for the loss and minimize the outage probability.

\begin{figure}
\begin{center}
\includegraphics[width=.8\linewidth, height=.5\linewidth]{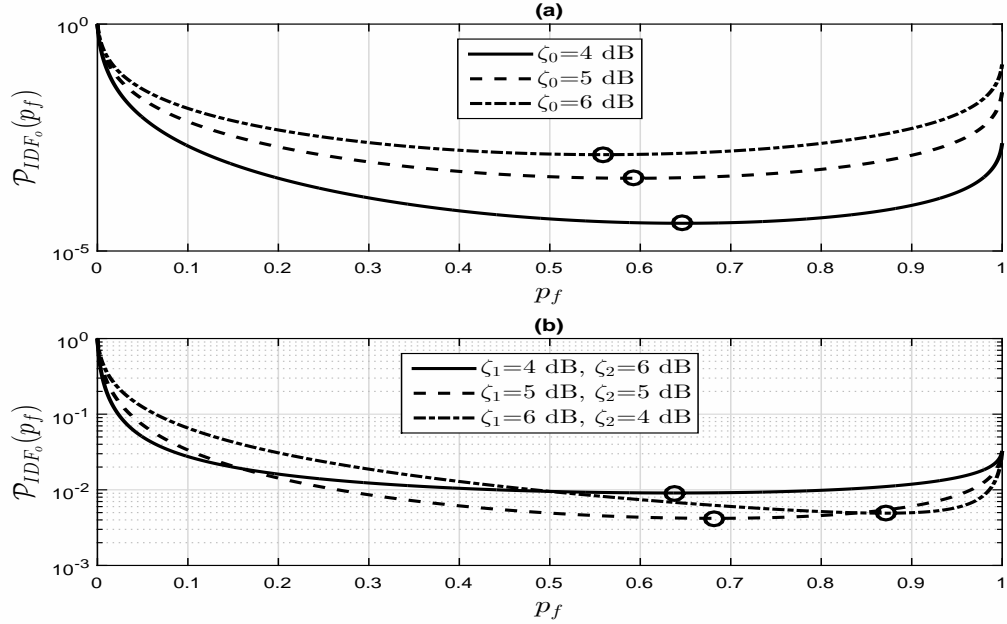}
\caption[]{Outage probability versus $p_f$ when imbalance in channel parameters with
$P_T=50$ dB, $d_{0}=200$ m, ${P}_{L}=80$ dB/km, $ \Lambda = 0.1$, 
$\eta = 10$, $R_{th}=2$, $d_f=0.5$, and $\Gamma_{1}=\Gamma_{th}$  when (a) $\xi_{1}=\xi_{2}=4$ dB and 
(b) $\xi_{0} =5$ dB.}
\label{FIG_POW2} 
\end{center}
\end{figure}

Fig. \ref{FIG_POW2}a depicts ${\cal P}_{\text{IDF}}(p_f)$ with respect to $p_f$ for different  $SD$ channel parameter at a given $SR$ and $RD$ channel quality, whereas Fig. \ref{FIG_POW2}b displays 
the same for different  $SR$ and $RD$ channel parameters at a given $SD$ channel quality.
The purpose of this figure is two fold: first, it shows how the direct link from $S$  to $D$ affects the power allocation when the relayed channel is balanced with equal $SR$ and $RD$ channel parameters, and secondly, it shows how the power allocation is affected when the relayed channel is imbalanced with unequal $SR$ and $RD$ channel quality at a given direct link channel quality. From  Fig. \ref{FIG_POW2}a, we can observe that as the direct 
link deteriorates due to the increase in channel parameters, 
the minimum point shifts towards right. This means that to minimize the outage, 
more and more power should be allocated to $S$ as the direct link quality worsens.

Fig. \ref{FIG_POW2}b {illustrates} that if the $SR$ channel quality is inferior compared to the  
$RD$ channel quality, the minimum point on outage probability is shifted towards right. This means that more power should be assigned 
to $S$ if the $SR$ channel quality is worse compared to the $RD$ channel quality. 
Thus, from the two cases, Figs. \ref{FIG_POW2}a and \ref{FIG_POW2}b, it can be concluded that more power should be allocated to $S$ to first secure the $SR$ transmission when compared to $RD$.
This is because in an incremental relaying scheme with $DF$ relays, 
if the first link can not maintain a proper rate of transmission, an outage is bound to happen even if the second link is better. 
By increasing the power at $S$, the direct transmission has better probability of success, and hence, the spectral efficiency increases.

Another point to be noticed from Fig. \ref{FIG_POW2} is that even if $SR$, $RD$, and $SD$ channel parameters are the same, the power allocation should not be equal, i.e., the optimum $p_f$ is not 0.5, rather higher than 0.5.
This is because for the same channel parameters even if $SR$ and $RD$ distances are the same, the path loss for $SD$ is always more than the other two path losses and must be compensated by providing more power to $S$. 
Thus, the power allocation must be designed as per the inter-node distances along with the channel parameters.

\section{Conclusion}
\label{sec_conclusion}
In this work, incremental relaying strategies have been introduced for the PLC system to improve spectral efficiency. Log-normal channel gain, Bernoulli-Gaussian impulsive noise, and distance-dependent attenuation are considered.
Closed-form expressions for the outage probability and fraction of times the relay is in use, along with approximate closed-form expressions 
for the end-to-end average channel capacity and the bit error rate for BPSK signaling are derived.
{Observations show} that at lower transmit power, the performance degrades as the required rate increases, while the converse happens at higher transmit power. 
The ISDF can achieve optimal spectral efficiency by choosing an appropriate threshold without compromising the performance. 
Power allocation studies reveal that power should be allocated according to the channel parameters and inter-node distances and not equally.
{We conclude that even if the relaying channels are identical, more than 50$\%$ of the total power should be allocated to the source to minimize the outage probability.}

\end{document}